\documentclass[preprint,showpacs,showkeys,preprintnumbers,amsmath,amssymb]{revtex4}

\usepackage{graphicx}
\usepackage{dcolumn}
\usepackage{bm}

\begin{document}

\preprint{}

\title{Fractal Fits to Riemann Zeros}

\author{Paul B. Slater}%
\email{slater@kitp.ucsb.edu}
\affiliation{%
ISBER, University of California, Santa Barbara, CA 93106\\
}%
\date{\today}

\begin{abstract}
Wu and Sprung ({\it Phys. Rev. E}, {\bf{48}}, 2595 (1993)) 
reproduced the first 500 
nontrivial Riemann zeros, using a one-dimensional local potential model. 
They concluded --- and similarly van Zyl and Hutchinson 
({\it Phys. Rev. E}, {\bf{67}}, 066211 (2003)) --- that 
the potential possesses a {\it fractal} 
structure of dimension $d =\frac{3}{2}$. 
We model 
the nonsmooth fluctuating part of the potential
by the alternating-sign sine series 
fractal of Berry and Lewis $A(x,\gamma)$. Setting $d=\frac{3}{2}$, 
we estimate the 
frequency parameter ($\gamma$), plus an overall scaling 
parameter ($\sigma$) we introduce.
We search for that pair of parameters ($\gamma,\sigma$) 
which {\it minimizes} the least-squares
fit $S_{n}(\gamma,\sigma)$ 
of 
the lowest $n$ eigenvalues --- obtained by solving the
one-dimensional stationary (non-fractal)
Schr\"odinger equation with the trial potential (smooth {\it plus} 
nonsmooth parts) --- to the 
lowest $n$ Riemann zeros for $n =25$. For the 
additional cases we study, $n=50$ and 75, 
we simply set $\sigma=1$. The fits obtained are 
compared to those gotten by  using just the {\it smooth}
part of the Wu-Sprung potential {\it without} any fractal
supplementation. Some limited improvement --- 5.7261 {\it vs.} 
6.39207 ($n=25$), 11.2672 {\it vs.} 11.7002 ($n=50$) and 
16.3119 {\it vs.} 16.6809 ($n=75$) --- is 
found in our (non-optimized, computationally-bound) 
search procedures.
The improvements are relatively strong in the vicinities of
$\gamma=3$ and (its {\it square}) 9.
Further, we extend the Wu-Sprung semiclassical framework to include
{\it higher-order} corrections from the Riemann-von Mangoldt formula 
(beyond the leading, dominant term)
into the smooth potential.
\newline
\newline
\end{abstract}


\maketitle

\section{Introduction}
In summarizing the results of their paper, ``Riemann zeros and a fractal potential'', Wu and Sprung stated that ``we have found analytically a one-dimensional local potential which generates the  smooth 
average level density obeyed by the Riemann zeros. We have then shown how any finite number of low lying Riemann zeros can be reproduced by introducing fluctuations on top of the potential.
The mystery of how a one-dimensional integrable system can produce a 
`chaotic' spectrum is resolved by adopting the concept of a fractal potential
which, in the infinite $N$ limit, would lead to the system having a dimension larger than one'' \cite[p. 2597]{wu}  (cf. \cite{khuri,tomiya,tomiya1,tomiya2,tomiya3,
dalidar,babinec}). (``Indeed \ldots finding an Hermitian operator whose
eigenvalues are [the Riemann zeros] may be impossible without introducing
chaotic systems'' \cite[p. 3]{khuri}.)

The Wu-Sprung potential $V$ --- which generates the smooth average
level density obeyed by the Riemann zeros --- satisfied 
{\it Abel's integral equation} 
\cite[eq. (6)]{wu}, and was
written {\it implicitly} as \cite[eq. (7)]{wu} (cf. \cite{diego} \cite[sec.~4]{jason}),
\begin{equation} \label{WSpotential}
x_{WS}(V)=\frac{1}{\pi} \Big( \sqrt{V-V_{0}} \ln \frac{V_{0}}{2 \pi e^2} +  \sqrt{V}
\ln \frac{\sqrt{V}+\sqrt{V-V_{0}}}{\sqrt{V}-\sqrt{V-V_{0}}} \Big).
\end{equation}
Here $V_{0}= 3.10073 \pi \approx 9.74123$.

Our objective is to reproduce, as best we can, 
the {\it fluctuations} on top of
the potential $V_{WS}(x)$, implicitly 
given by (\ref{WSpotential}), so that the application of the 
Schr\"odinger equation to the so-amended (smooth {\it plus} fractal) 
potential would 
yield the Riemann zeros {\it themselves}.
For our exploratory purposes, we adopt 
(being a particular case of a {\it deterministic} 
Weierstrass-Mandelbrot [WM] fractal 
function) the alternating-sign sine series of
Berry and Lewis \cite[eq. (5)]{berry},
\begin{equation} \label{alternating}
A(x,\gamma) =\Sigma_{m=-\infty}^{\infty} 
\frac{(-1)^m \sin{\gamma^m x}}{\gamma^{(2-d) m}}, \hspace{1in}                 (1 <d  < 2, 1 <  \gamma)                    .
\end{equation}
Here, $d$ is the fractal dimension, which --- following 
the  box-counting argument of 
Wu and Sprung \cite{wu}
(cf. \cite{vanZyl}) --- we take to be $\frac{3}{2}$.
We have, in this $d=\frac{3}{2}$ Berry-Lewis context,  a specific case, 
\begin{equation} \label{AffineScaling}
A(\gamma x,\gamma)= - \gamma^{\frac{1}{2}}  A(x,\gamma),
\end{equation}
of the  ``affine scaling law'' \cite[eq. (3)]{berry}.
We also scale --- in the first ($n=25$) of our three sets of analyses 
($n=25,50,75$) --- $A(x,\gamma)$ by a parameter $\sigma$, 
where $n$ is the number of the lowest Riemann zeros we aspire to fit 
(sec.~\ref{secn25}).
For the cases $n=50$ (sec.~\ref{secn50}) and 75 
(sec.~\ref{secn75}), we will simply set $\sigma=1$.
In sec.~\ref{extend}, we demonstrate how to incorporate
more terms of the Riemann-von Mangoldt formula \cite{karatsuba} for the 
cumulated number of Riemann zeros 
than Wu-Sprung themselves did, using a semiclassical 
argument, in deriving $x_{WS}(V)$.
(It remains, however, to numerically implement these last findings.)
\section{Analyses}
\subsection{$n=25$} \label{secn25}

We proceed, to begin, trying 
to fit the first twenty-five Riemann zeros 
by finding distinguished values of the two parameters ($\gamma$ and $\sigma$).
We randomly generate trial values $1 \leq \gamma \leq 10$ and
$0 \leq \sigma <10$. (Numerically-speaking, we truncate the summation 
in (\ref{alternating}) by summing from $m=-30$ to $m=30$ (cf. \cite[App.]{berry}). We have not yet gauged the sensitivity of the 
various results in this paper
to this choice of
cutoff --- nor to the setting $d=\frac{3}{2}$ {\it nor} 
further to the specific 
measure of fit (sum-of-square deviations) employed --- though it would 
certainly be of interest to do so for any or all of them.)

\begin{figure} 
\includegraphics{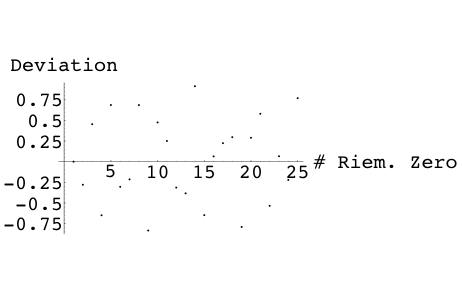}
\caption{\label{fig:RiemDev}Deviations from the first twenty-five Riemann zeros of the first twenty-five eigenvalues obtained by solving the Schr\"odinger
equation using the (smooth/non-fractal) 
Wu-Sprung potential (\ref{WSpotential}). 
The sum-of-squares of these deviations is 6.39207, while the 
sum-of-squares of these Riemann zeros is much larger, 92569.63.}
\end{figure}

If we use the {\it smooth} potential given by 
(\ref{WSpotential}) itself --- {\it without}
any fractal supplementation --- the sum-of-squares deviation of the
first twenty-five eigenvalues yielded by application of the 
one-dimensional stationary 
Schr\"odinger equation from the first twenty-five Riemann zeros is
6.23907 (Fig.~\ref{fig:RiemDev}). (This is only 0.0069 percent of 
the total [non-fitted] 
sum of squares of the zeros themselves, that is, 92569.63, 
so one might aver that the semiclassically-based 
smooth Wu-Sprung potential is notably successful
in well-approximating the Riemann zeros. It is, of course, our objective
here to {\it reduce} this small percentage even further. Let us also note
that a referee suggested that the scatter in Figs.~ \ref{fig:RiemDev} 
\ref{fig:RiemDev2} 
and \ref{fig:50zeros} 
might be reduced if the modulus of the scatter were to be
plotted.)

\begin{figure}
\includegraphics{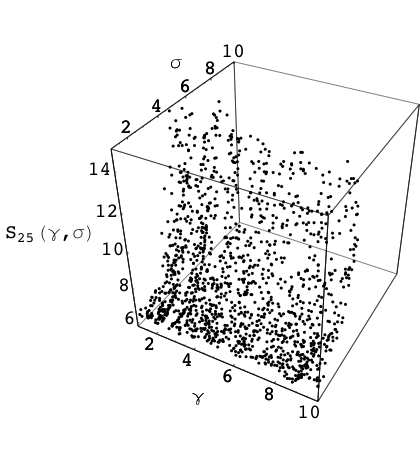}
\caption{\label{fig:Scatter3D}Scatterplot showing 
the sum-of-squares goodness-of-fit statistic $S_{25}(\gamma,\sigma)$ 
as a function of the frequency parameter ($\gamma$) and the
scaling parameter ($\sigma$). Those 1,833 points of the 4,007 sampled 
for which 
$S_{25}(\gamma,\sigma)<15$ are included.}
\end{figure}

We 
randomly generated 4,007 pairs of ($\gamma,\sigma$) from the indicated
ranges, and solved for {\it each} such pair,  
the Schr\"odinger equation with the smooth potential
{\it plus} the fractal $\sigma A(x,\gamma)$ with the corresponding 
choice of $\gamma$ and $\sigma$. 
Further, we calculated the 
corresponding sum-of-squares ($S_{25}(\gamma,\sigma)$) deviations
from the first twenty-five Riemann zeros. We obtained a range of values
$S_{25}(\gamma,\sigma) \in [5.7261,166.075]$. 
In Fig.~\ref{fig:Scatter3D}, we display the results 
for those 1,833 pairs of the 4,007 that yielded $S_{25}(\gamma,\sigma)<15$.

\begin{figure}
\includegraphics{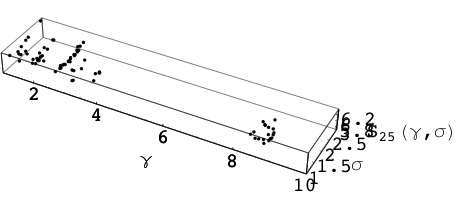}
\caption{\label{fig:Scatter3Da}Truncation of the
previous scatterplot to those 210 (fit-improving)
pairs yielding $S_{25}(\gamma,\sigma) <6.39207$. For {\it all} such pairs
$\sigma <2.86475$.}
\end{figure}

\begin{figure}
\includegraphics{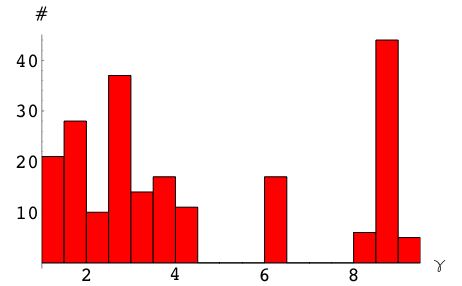}
\caption{\label{fig:Histogram1}Histogram for the frequency parameter
$\gamma$ corresponding to those 210 pairs $(\gamma,\sigma)$ for which
$S_{25}(\gamma,\sigma) < 6.39207$, the value obtained from solving 
the Schr\"odinger equation using the smooth
{\it unsupplemented} Wu-Sprung potential. The classification bins 
for $\gamma \in [1,9.5]$ are of
length $\frac{1}{2}$.}
\end{figure}

Additionally, in Fig.~\ref{fig:Scatter3Da} we show the results for those 210 pairs yielding
$S_{25}(\gamma,\sigma) < 6.39207$ --- that is, those pairs which yield results {\it superior} (numerically inferior, that is) 
to those obtained with the original smooth {\it unsupplemented} potential
(\ref{WSpotential}).
Fig.~\ref{fig:Histogram1} is a histogram of the values of $\gamma$ --- having 
been uniformly sampled from [1,10] --- occurring in 
these 210 pairs. (There were {\it no} $\gamma$'s recorded for a 
number of bins, including [9.5,10].)
The two topmost peaks (corresponding to the classification bins 
$\gamma \in [2.5,3]$ and [8.5,9]) 
may possibly reflect  (cf. (\ref{AffineScaling})) an assertion of
Berry and Lewis, made (using $t$ for what we denote 
by $x$) in regard to what they termed deterministic 
Weierstrass-Mandelbrot fractal functions ($W(t)$). They note that
``the whole function $W$ can be reconstructed from its values in the
range $t_{0} \leq t \leq \gamma t_{0}$; for example, $W$ in the ranges
$\gamma t_{0} \leq t \leq \gamma^{2} t_{0}$ and 
$\gamma^{-1} t_{0} \leq t \leq t_{0}$ are magnified and diminished
versions, respectively, of $W$ in the range $t_{0} \leq t \leq \gamma
 t_{0} \ldots $
The repetition and resolution of features at $t_{0}, \gamma t_{0}, 
\gamma^2 t_{0}$ etc, is again obvious'' \cite[p. 461]{berry}.
(Similar $\gamma$-histograms --- Figs.~\ref{fig:Histogram50gamma} and
\ref{fig:Histogram75gamma} --- will be obtained for the $n=50$ and 75 
analyses further below.)

\begin{figure}
\includegraphics{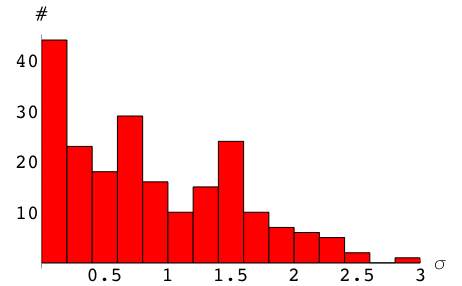}
\caption{\label{fig:Histogram2}Histogram for the scaling parameter
$\sigma$ corresponding to those 210 pairs $(\gamma,\sigma)$ for which
$S_{25}(\gamma,\sigma) < 6.39207$}
\end{figure}

\begin{figure}
\includegraphics{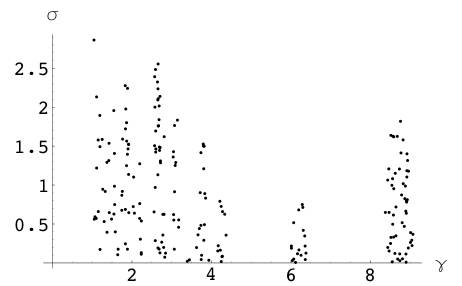}
\caption{\label{fig:correlation}Plot of those pairs ($\gamma,\sigma$) 
for which $S_{25}(\gamma,\sigma) < 6.39207$. The correlation coefficient
is -0.264263.}
\end{figure}

\begin{figure}
\includegraphics{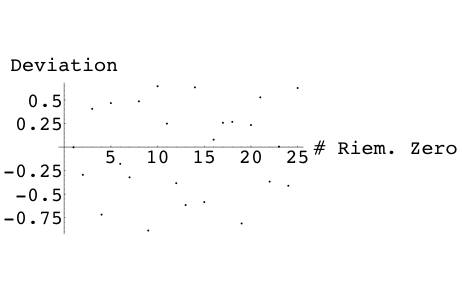}
\caption{\label{fig:RiemDev2}Deviations from the first twenty-five 
Riemann zeros of the first twenty-five
eigenvalues obtained by solving the Schr\"odinger
equation using the (non-fractal) Wu-Sprung potential (\ref{WSpotential})
{\it supplemented} by the scaled Berry-Lewis fractal function 
$1.95798 A(x,1.54523)$.
The sum-of-squares of these deviations is 5.7261, reduced from 6.39207 
for the non-supplemented smooth potential (Fig.~\ref{fig:RiemDev}).}
\end{figure}

In Fig.~\ref{fig:Histogram2} we have an analogous histogram based
on the scaling parameter $\sigma$. (For {\it no} value 
$\sigma> 2.86475$, though we sample {\it uniformly} from [0,10],
do we obtain an improvement by using $\sigma A(x,\gamma)$.)
In Fig.~\ref{fig:correlation} we have --- for these same 210 pairs --- a 
plot of $\gamma$ {\it vs.} $\sigma$. (The associated correlation 
coefficient is negative, that is, -0.275492.)

\begin{figure}
\includegraphics{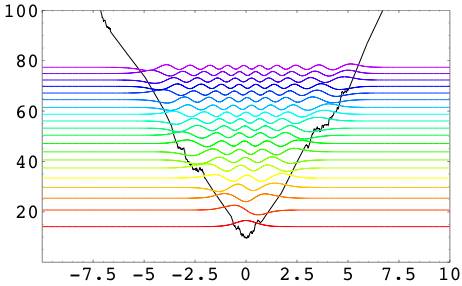}
\caption{\label{fig:FractalPotential25}Eigenfunctions drawn at the
eigenvalues obtained by solving the Schr\"odinger equation 
for the potential which yields the minimum $S_{25}(1.54523,1.95798)=5.7261$ of
the 4,007 pairs sampled}
\end{figure}

The {\it minimum} 
over all the 4,007 pairs was $S_{25}(1.54523,1.95798)$ = 5.7261.
(The next smaller value was $S_{25}(1.15274,1.57931)= 5.81754$. 
All other sum-of-squares deviations {\it exceeded} 6.0.)
In Fig.~\ref{fig:RiemDev2} (cf. Fig.~\ref{fig:RiemDev}) we show 
the deviations of the predicted eigenvalues at this point (1.15274,1.57931) 
from the 
corresponding Riemann zeros.
In Fig.~\ref{fig:FractalPotential25} we display, for this same minimizing
point, the corresponding twenty-five eigenfunctions drawn at the 
twenty-five eigenvalues for
the associated (smooth plus fractal) potential.

\subsection{$n=50$} \label{secn50}
\begin{figure}
\includegraphics{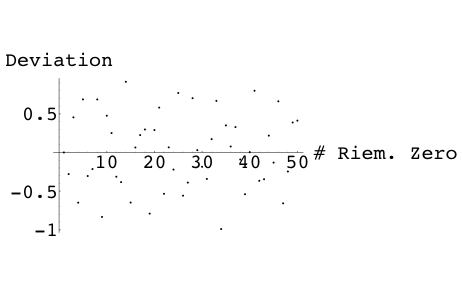}
\caption{\label{fig:50zeros}Deviations 
from the first {\it fifty} Riemann zeros of the first {\it fifty}
eigenvalues obtained by solving the Schr\"odinger
equation using the (non-fractal) Wu-Sprung potential (\ref{WSpotential}).
The sum-of-squares of these deviations is 11.7002.}
\end{figure}

Now, we present 
in Fig.~\ref{fig:50zeros} the extension of  Fig.~\ref{fig:RiemDev} 
from the first twenty-five to 
the first {\it fifty} Riemann
zeros. 
(It is considerably more demanding to solve the Schr\"odinger equation 
for this increased $n$.)
The associated sum-of-squares is 11.7002, which is but 0.00261 percent 
of the total sum-of-squares, 448704.56, 
of the first fifty Riemann zeros themselves.

\begin{figure}
\includegraphics{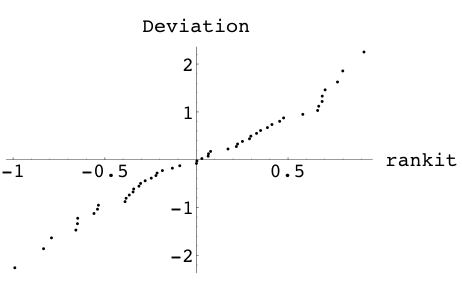}
\caption{\label{fig:rankit}A ``rankit'' plot to assess 
the possible {\it normality} of the
distribution of deviations of the fifty eigenvalues from the Schr\"odinger
equation solution using the smooth Wu-Sprung potential from the Riemann
zeros themselves. For normally-distributed observations, one expects a
{\it straight} line.}
\end{figure}

In Fig.~\ref{fig:rankit} we show the results of a ``rankit'' test 
(described in the {\it Wikipedia} on-line encyclopaedia) for 
the {\it normality} of the distribution of deviations in Fig.~\ref{fig:50zeros}. If the observations do come from a normal/Gaussian distribution, we expect
the plot (necessarily non-decreasing in any case, Gaussian or otherwise) 
to be a {\it straight} line. So, there appears to be some deviations from
strict normality, particularly in the tails of the distribution 
(suggesting that perhaps sums-of-{\it squares} might not be the most
robust measure of deviation to be employed for evaluation of 
fits --- though it is certainly 
the most conventional and familiar measure).

\begin{figure}
\includegraphics{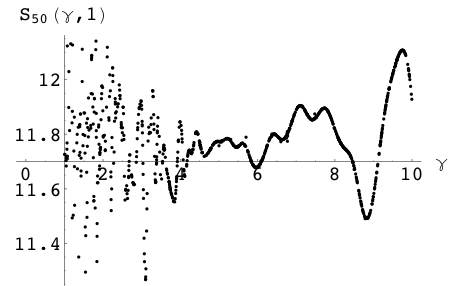}
\caption{\label{fig:MoreZeros}Scatterplot showing the sum-of-squares
goodness-of-fit statistic $S_{50}(\gamma,1)$, based on 1,013
randomly chosen points $0 \leq \gamma \leq 10$. 
Points below the $\gamma$-axis
correspond to {\it improvements} in fit.}
\end{figure}

We repeated for the case $n=50$, 
the form of analysis conducted 
for $n=25$ (sec.~\ref{secn25}), 
but now omitting 
(in the interest of interpretational simplicity) 
the scaling parameter
($\sigma$) --- effectively setting it to unity.
Based on 1,013 random choices of $\gamma$ lying in $[1,10]$, we obtained
Fig.~\ref{fig:MoreZeros}.
The {\it minimum} achieved was $S_{50}(3.10007,1) = 11.2672$,
while without any fractal supplementation at all, the
(larger) value of 11.7002 was
obtained.
(As a matter of curiosity, 
we found that $S_{50}(1,g_{Au})=11.7641 \geq 11.7002$, where
$g_{Au}=\frac{\sqrt{5}+1}{2} \approx 1.61803$ is the {\it golden mean} 
\cite{livio}. Interpolation by third-degree polynomials 
of the points in Fig.~\ref{fig:MoreZeros} 
suggests that the actual minimum of 11.2556 would be achieved 
for $\gamma=3.092$.) The penultimate minimum, 
$S_{50}(3.10051,1)=11.2685$, was nearby.
Again (cf. Fig.~\ref{fig:Histogram1}), 
we appear to discern a manifestation of relative minima in the vicinity of both
$\gamma=3$ and (its {\it square}) 9. (Also around $\gamma=6$, as in 
Fig.~\ref{fig:Histogram1} too. The 
associated relative minimum of 11.678 occurs at
$\gamma=5.96158$.)
The relative minimum (11.4906) in the vicinity of $\gamma=9$ was 
attained at $\gamma=8.78602$, so, at that point, one has
$\sqrt{\gamma}= 2.96412$.
For the smaller values of $\gamma$, the plot is very scattered, but for 
the higher ones, a well-defined curve emerges. (The overall {\it maximum} is
12.1412 at $\gamma=1.8165$, 
while the maximum in the upper range $[8,10]$ is 
12.108 at $\gamma=9.7516$.)
We applied (third-order) interpolation to the data in 
Fig.~\ref{fig:MoreZeros} and
obtained for the most salient domains of {\it improved} fit,
$\gamma \in [3.03205,3.1537]$ (containing our overall minimum) 
and [3.33583,3.38488], [5.85954,6.08565] and [8.45962,9.12272] 
(containing our three distinct relative minima [Fig.~\ref{fig:MoreZeros}]).

\begin{figure}
\includegraphics{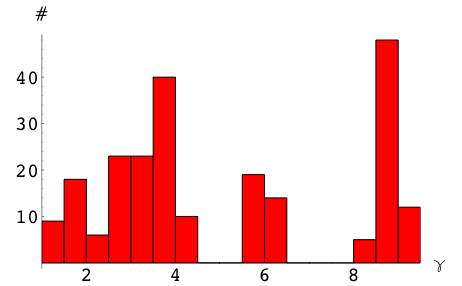}
\caption{\label{fig:Histogram50gamma}Histogram 
corresponding to those 219 values of 
the frequency parameter $\gamma$ for which
$S_{50}(\gamma,1) < 11.7002$, the value obtained from solving
the Schr\"odinger equation using the smooth
{\it unsupplemented} Wu-Sprung potential. The classification bins 
for $\gamma \in [1,9.5]$ are of
length $\frac{1}{2}$.}
\end{figure}

\begin{figure}
\includegraphics{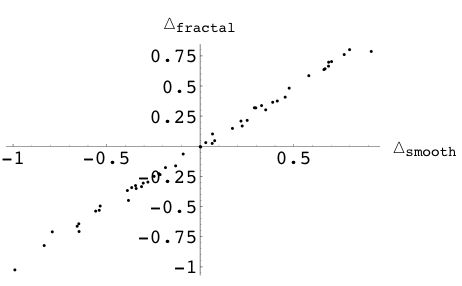}
\caption{\label{fig:compareDev}The predicted eigenvalues minus the first
fifty Riemann zeros: the horizontal axis based on the smooth potential and
the vertical axis based on the sum-of-squares minimizing fit}
\end{figure}

\begin{figure}
\includegraphics{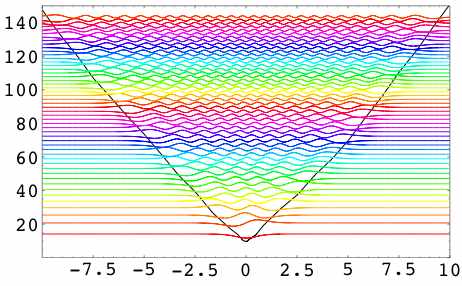}
\caption{\label{fig:FractalPotential50}Eigenfunctions drawn at the
eigenvalues obtained by solving the Schr\"odinger equation
for the potential which yields the minimum $S_{50}(3.10007,1)=11.2672$ of
the 1,013 points sampled}
\end{figure}

In Fig.~\ref{fig:Histogram50gamma} (cf. Fig.~\ref{fig:Histogram1}) 
we show the histogram based on those (fit-improving) 219 of the 1,013 
values of $\gamma$ for which $S_{50}(\gamma,1)<11.7002$.
In Fig.~\ref{fig:compareDev} we plot the deviations 
($\Delta_{smooth}$) from the first 
fifty Riemann zeros obtained using the smooth potential against those 
deviations ($\Delta_{fractal}$) using
the supplemented potential which minimizes the fit, obtaining basically
a {\it linear} relationship.
In Fig.~\ref{fig:FractalPotential50}, we have the counterpart
of Fig.~\ref{fig:FractalPotential25}, for the case $n=50$.
\subsection{$n=75$}\label{secn75}
\begin{figure}
\includegraphics{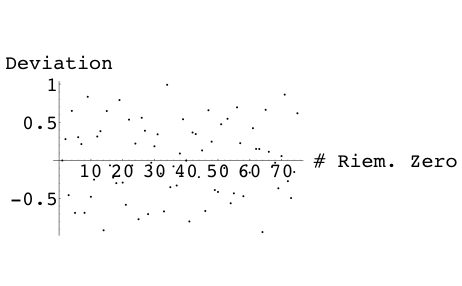}
\caption{\label{fig:75zeros}Deviations
from the first {\it seventy-five} Riemann zeros of the first {\it 
seventy-five}
eigenvalues obtained by solving the Schr\"odinger
equation using the (non-fractal) Wu-Sprung potential (\ref{WSpotential}).
The sum-of-squares of these deviations is 16.6809.}
\end{figure}

\begin{figure}
\includegraphics{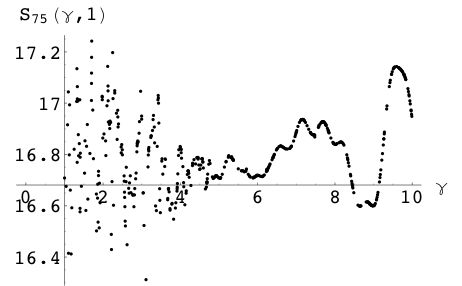}
\caption{\label{fig:n75}Scatterplot showing the sum-of-squares
goodness-of-fit statistic $S_{75}(\gamma,1)$, based on 853
randomly chosen points $0 \leq \gamma \leq 10$.
Points --- of which there are 172 --- below the $\gamma$-axis
correspond to {\it improvements} in fit.}
\end{figure}

\begin{figure}
\includegraphics{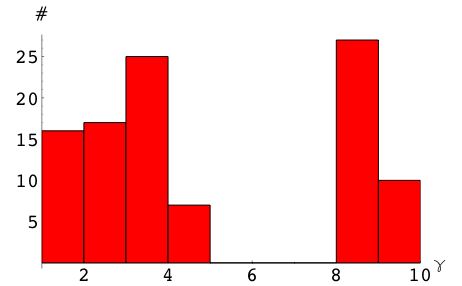}
\caption{\label{fig:Histogram75gamma}Histogram
corresponding to those 172 values of
the frequency parameter $\gamma$ for which
$S_{75}(\gamma,1) < 16.6809$, the value obtained from solving
the Schr\"odinger equation using the smooth
{\it unsupplemented} Wu-Sprung potential. The classification bins
for $\gamma \in [1,10]$ are of
length $1$.}
\end{figure}

Fig.~\ref{fig:75zeros} is the further extension of Figs.~\ref{fig:RiemDev}
and \ref{fig:50zeros} to the $n=75$ case. 
The associated sum-of-squares is 16.6809, which is but 
0.0014308 {\it percent} of the total sum-of-squares of the 
first seventy-five Riemann zeros, $1.1658469 \cdot 10^6$.
We generated 853 values of the frequency parameter $\gamma$, uniformly
sampling from the interval [1,10]. The best fit of 16.3119 was achieved
for $\gamma=3.1106$. 
(The penultimate minimum of 16.3427 was nearby at $\gamma=3.11632$, 
while the overall maximum [worse fit] was 19.27 at
$\gamma=2.47689$.)
In Fig.~\ref{fig:n75} we plot these 853 values as a function
of $\gamma$.
(In the vicinity of $\gamma=9$, the relative minimum 
of 16.5983 is achieved at
$\gamma=8.64589 \approx 2.94039^2$.) 
In Fig.~\ref{fig:Histogram75gamma} we show the corresponding
histogram.

We continue to add randomly generated points to this $n=75$ analysis,
and may possibly undertake an $n=100$ study too.
\section{Higher-order corrections to the Wu-Sprung potential} \label{extend}
In their semiclassical analysis, Wu and Sprung \cite{wu} took into account
only the leading term of the Riemann-von Mangoldt formula \cite{karatsuba}, 
but nevertheless, as our results indicate, 
doing so is able to yield, {\it via} the Schr\"odinger equation,
eigenvalues that 
quite closely approximate the Riemann zeros themselves. It might 
prove beneficial, in trying to account for the (relatively small) residual 
variation --- especially since we have been working in a 
non-asymptotic regime --- to extend the
Wu-Sprung approach by incorporating the higher-order non-oscillatory terms
of the Riemann-von Mangoldt formula too 
(cf. \cite{weibert} \cite[eq. (5)]{main}).

The (smooth) leading term --- of the Riemann-von Mangoldt formula for 
the number of zeros below $E$ --- employed 
by Wu and Sprung took the form
\begin{equation} \label{leadingterm1}
N(E) =\frac{E}{2 \pi} \log{\frac{E}{2 \pi e}} +\frac{7}{8}.
\end{equation}
(This formula pertains to the important 
[Connes/Berry-Keating---absorption/emission spectrum] ``sign'' problem 
\cite[eqs. (6), (11)]{sierra}.) 
The remaining part (having {\it both} smooth and nonsmooth components) 
{\it not} utilized
by Wu and Sprung is expressible as \cite[p. 436]{karatsuba}
\begin{equation}
S(E)+\frac{1}{\pi} \delta (t),
\end{equation}
where
\begin{equation}
S(E) =\frac{1}{2} \arg \zeta (\frac{1}{2} + i E),
\end{equation}
is the {\it argument} function and 
$\zeta(s)$ is the Riemann zeta function. Further,
\begin{equation}
\delta(E) = \frac{E}{4} \log{ \Big( 1+\frac{1}{4 E^2} \Big)} +\frac{1}{4} 
\arctan{\frac{1}{2 E}}- \frac{E}{2} \int_{0}^{+ \infty} 
\frac{\rho (u) d u}{(u+1/4)^2 +(E/2)^2}
\end{equation}
and $\rho (u) = 1/2-\{u\}$. (Here $\{u\}$ is the fractional part of $u$.)
The function $S(E)$ is itself strongly oscillatory (changing sign an {\it 
infinite} number of times).

Incorporating the two additional (higher-order --- $O(E^{-2})$ and $O(E^{-3}$), respectively) 
non-oscillatory (monotonically decreasing)
terms of the Riemann-von Mangoldt formula \cite{karatsuba}, one could seek to 
find  the potential based on
\begin{equation} \label{level}
N(E) + \frac{E}{4} \log{ \Big( 1+\frac{1}{4 E^2} \Big)} +\frac{1}{4}
\arctan{\frac{1}{2 E}}.
\end{equation}
The corresponding potential can, in fact, be 
straightforwardly constructed (in implicit form), 
using the formula for Abel's 
integral equation \cite{gorenflo}, following the procedure outlined in
\cite{wu}. It takes the form
\begin{equation} \label{Corrections}
x_{WS_{H}}(x)= \frac{2 (1+\pi ) \sqrt{V} \tanh
   ^{-1}\left(\frac{\sqrt{V-V_0}}{\sqrt{V}}\right)}{\pi
   ^2}-\frac{2 (1+\pi ) \sqrt{V-V_0}}{\pi ^2}+\frac{2
   \sqrt{V-V_0}}{\pi ^2}
\end{equation}
\begin{displaymath}
-\frac{V \tanh
   ^{-1}\left(\frac{\sqrt{V-V_0}}{\sqrt{V-\frac{i}{2}}}\right)}{\pi ^2 \sqrt{V-\frac{i}{2}}}-\frac{V \tanh
   ^{-1}\left(\frac{\sqrt{V-V_0}}{\sqrt{V+\frac{i}{2}}}\right)}{\pi ^2 \sqrt{V+\frac{i}{2}}}
\end{displaymath}
\begin{displaymath}
-\frac{i \left(\frac{\tan ^{-1}\left(\frac{\sqrt{2}
   \sqrt{V-V_0}}{\sqrt{2 V_0-i}}\right)}{\sqrt{2
   V_0-i}}-\frac{\tan ^{-1}\left(\frac{\sqrt{2}
   \sqrt{V-V_0}}{\sqrt{2 V_0+i}}\right)}{\sqrt{2
   V_0+i}}\right)}{\sqrt{2} \pi ^2}
\end{displaymath}
\begin{displaymath}
-\frac{\left(\log \left(1+\frac{1}{4 V_0^2}\right)+\pi 
   \left(\log \left(4 \pi ^2\right)-2 \log
   \left(V_0\right)\right)\right) \sqrt{V-V_0}}{2 \pi ^2}
\end{displaymath}
\begin{displaymath}
-\frac{i \log \left(\frac{(1-i) V_0+\sqrt{V}}{\left((1+i)
   \sqrt{V}+i\right) \sqrt{2 i-4 V_0}}+\frac{2 i
   \sqrt{V-V_0}}{4 \sqrt{V}+(2+2 i)}\right)}{2 \pi ^2
   \sqrt{2 i-4 V_0}}
\end{displaymath}
\begin{displaymath}
+\frac{i \log \left(\frac{(-1+i) V_0-i
   \sqrt{V}}{\left((1+i) \sqrt{V}-1\right) \sqrt{-4 V_0-2
   i}}-\frac{2 i \sqrt{V-V_0}}{4 \sqrt{V}-(2-2
   i)}\right)}{2 \pi ^2 \sqrt{-4 V_0-2 i}}
\end{displaymath}
\begin{displaymath}
-\frac{i \log \left(\frac{(1-i) \left(\sqrt{i-2 V_0}
   \sqrt{V-V_0}-\sqrt{2} V_0\right)+\sqrt{2}
   \sqrt{V}}{\left(2 i-(2+2 i) \sqrt{V}\right) \sqrt{i-2
   V_0}}\right)}{2 \pi ^2 \sqrt{2 i-4 V_0}}
\end{displaymath}
\begin{displaymath}
+\frac{i \log \left(\frac{i \left((1+i)
   V_0+\sqrt{V}\right)}{\left((1+i) \sqrt{V}+1\right)
   \sqrt{-4 V_0-2 i}}+\frac{(1-i) \sqrt{V-V_0}}{(2+2 i)
   \sqrt{V}+2}\right)}{2 \pi ^2 \sqrt{-4 V_0-2 i}}.
\end{displaymath}

\begin{figure}[ht]
\includegraphics{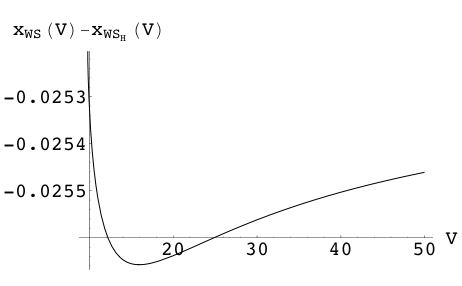}
\caption{\label{fig:DiffPot1}The original 
(implicitly expressed) Wu-Sprung potential (\ref{WSpotential}) minus
the one (\ref{Corrections}) with higher-order 
Riemann-von Mangoldt corrections}
\end{figure}

\begin{figure}[ht]
\includegraphics{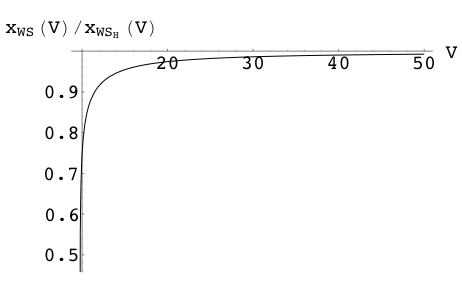}
\caption{\label{fig:DiffPot2}The original (implicitly expressed) 
Wu-Sprung potential 
(\ref{WSpotential}) divided by
the one (\ref{Corrections}) with higher-order 
Riemann-von Mangoldt corrections}
\end{figure}

We have not so far been able to numerically invert $x_{WS_{H}}(V)$, but in 
Figs.~\ref{fig:DiffPot1} and \ref{fig:DiffPot2} we show the difference
$x_{WS}(V)-x_{WS_{H}}(V)$ and the ratio $\frac{x_{WS}(V)}{x_{WS_{H}}(V)}$, respectively.
(As in the original 
Wu-Sprung analysis \cite{wu}, we set $V_{0}=9.74123$ in both cases.)

\section{Concluding Remarks}
It might be possibly worthwhile exploring, in our context, 
the use of fractals other 
than the specific Berry-Lewis alternating-sign sine series (\ref{alternating}) 
(cf. \cite[Chap. 12]{tricot} \cite{carlos1,slaterRiemann}). 
(It appears that
$A(x,\gamma)$ {\it increases} with
$x$, while the Wu-Sprung 
plots of the fluctuations of the fractal potential from the 
smooth one \cite[Fig. 2]{wu} do {\it not} seem to exhibit such a growth
phenomenon (cf. \cite[Fig. 12.13]{tricot}). However, 
certain quite preliminary efforts
of ours to explore along such lines have encountered some so-far
not well-understood numerical difficulties.) 
Along with the alternating-sign sine series ($A(x,\gamma$))  
we have employed, Berry and 
Lewis too analyzed a certain companion 
(also deterministic Weierstrass-Mandelbrot)
cosine series $C(x,\gamma)$. 
However, this function can only
assume nonnegative values, so --- unless translated  --- it does not 
provide an immediately suitable model of the Wu-Sprung 
({\it both} negative and positive) fluctuations.

In the asymptotic 
limit, the Wu-Sprung potential 
$V_{WS}(x)$  behaves as \cite[eq. (9)]{diego},
\begin{equation} \label{DD}
V_{WS_{asymp}}(x) = \frac{\pi^2 x^2}{4} \Big[ \text{LW} 
\Big( \sqrt{\frac{\pi}{2}} \frac{|x|}{e} \Big) \Big]^{-2}, |x| \rightarrow \infty,
\end{equation}
where LW($\cdot$) represents the Lambert-W function \cite{knuth}.
It would be interesting to explore the question of whether the Schr\"odinger
equation can be {\it exactly} 
solved with (\ref{DD}) as a potential (cf. \cite{williams}).
(In regard to this matter, M. Trott commented that 
``I would guess it is enough to look for the asymptotic region.
If an analytic solution exists there, one might be able to
'guess' the correct one in terms of Lambert-W.
If one can't find an exact solution asymptotically, then probably
no exact one in Lambert-W does exist either''.)
If so, then with the use of {\it perturbation theory} \cite{kato}, 
one might possibly be 
able to expedite the computational procedure we have employed above 
(in which we have solved the Schr\"odinger 
equation --- a demanding task --- {\it ab initio} for
{\it each} new set of random parameters).

Questions pertaining to the nature of the eigen{\it functions} obtained and
of the importance of tunneling  contributions in the fractal potential 
remain to be addressed (cf. Figs.~\ref{fig:FractalPotential25} and 
\ref{fig:FractalPotential50}).

Castro and Mahecha --- in their proposed {\it supersymmetric} model of 
the Riemann zeros --- suggested the use of a ``{\it fractal} 
[emphasis added]  SUSY QM equation 
instead of factoring the {\it ordinary} [emphasis added]  
Schr\"odinger equation studied
by Wu-Sprung'' \cite[p. 788]{carlos1} 
(cf. \cite{slaterRiemann,laskin,ben-adda}). 
They further 
proposed the use of a fully 
general Weierstrass-Mandelbrot fractal to model
fluctuations in the supersymmetric potential. 
The alternating-sign sine series fractal (\ref{alternating}) 
we have employed in this study
is a {\it specific} 
instance of the WM-fractal \cite{berry}, involving far {\it fewer} 
parameters --- which is largely why we have employed it here 
in our exploratory numerical analyses. 
(References to further related work of Castro can be found at the 
number theory and physics archive website 
http://secamlocal.ex.ac.uk/people/staff/mrwatkin/zeta/physics.htm.)

The analytical approach of Wu and Sprung \cite{wu}, which has been the
basis of our study here incorporates the familiar (``Berry-Keating'' 
\cite{berrykeating}) 
form of density-of-states $N(E)$ given in 
(\ref{leadingterm1}). However, in Connes's adelic
(absorption spectrum) approach \cite{connes} the density-of-states takes
another (cutoff ($\Lambda$)-dependent) form \cite[eq. (11)]{sierra},
\begin{equation}
N_{Connes}(E) = \frac{E}{2 \pi} \log{\Lambda^2} -\frac{E}{2 \pi} 
\Big(\log{\frac{E}{2 \pi}} -1\Big).
\end{equation}
The question, then, arises of whether or not the Wu-Sprung framework can 
be meaningfully/differently  recast in the Connes adelic setting.
(Berry has been quoted to the effect that ``we think this is a matter 
of formalism and everything you can write as an absorption you can write as 
an emission spectrum if you manipulate things the right way'' 
\cite[p. 247]{sabbagh}. It has been proposed that the approach of Connes
can be implemented using pseudo-Hermitian Hamiltonians for which the 
eigenvectors are null \cite{ahmed}. Shudo {\it et al} 
\cite{shudo} measured 
{\it absorption} spectra and investigated the distribution of eigenfrequencies
in ``L-shaped'' resonators.)

In the context of {\it fractal} strings and sprays, the classical 
Riemann zeta function ``can be viewed as the spectral zeta function of
the unit interval'' \cite[p. 2]{michel}.

In their study of
the Weierstrass-Mandelbrot fractal function, upon which we have 
strongly relied,
Berry and Lewis found, after some delicate analysis,
that an attractive inverse-square potential
generates the ``Weierstrass spectrum'' $\gamma^n$ \cite[sec. 5]{berry}. 
They further 
speculated that
there might be no other such potential.
\begin{acknowledgments}
I would like 
to express gratitude to the Kavli Institute for Theoretical
Physics (KITP)
for computational support, and
to Carlos Castro and Michael Trott for their sustained interest and 
expert advice. M. Trott first implemented the analysis underlying
Figs.~\ref{fig:RiemDev} and \ref{fig:FractalPotential25} and 
furnished the Mathematica program for solving
the Schr\"odinger equation that was used repeatedly above.
\end{acknowledgments}

\bibliography{Riemann3}

\end{document}